\documentclass[psfig,useAMS,usenatbib]{mn2e}
\usepackage{times}
\usepackage{epsfig}
\usepackage{amsmath}
\usepackage{natbib}
\usepackage{longtable}
\usepackage{float}
\usepackage{multirow}

\def\apj {ApJ}
\def\apjl {ApJL}
\def\apjs {ApJS}
\def\aj {AJ}

\def\aap {A\&A}
\def\mnras {MNRAS}
\def\pasp {PASP}

\def\etal{{\rm et al. }}

\def\kpc{{\ h^{-1} \ \rm kpc}}
\def\kms{{\ \rm km \ s^{-1}}}

\title{Interacting galaxies: co-rotating and counter-rotating systems with tidal tails}

\author[Mesa et al.]{ Valeria Mesa$^{1,2}$,
 Fernanda Duplancic$^{1,3}$,
 Sol Alonso$^{1,3}$,
 Georgina Coldwell$^{1,3}$ \& \newauthor
 Diego G. Lambas$^{1,4}$\\
 $^{1}$ Consejo de Investigaciones Cient\'\i ficas y T\'ecnicas (CONICET), Avenida Rivadavia 
1917, C1033AAJ, Buenos Aires, Argentina \\
$^{2}$ Instituto Argentino de Nivolog\'{i}a Glaciolog\'{i}a y Ciencias Ambientales (IANIGLA), Parque Gral San Mart\'{i}n, CC 330, CP 5500, Mendoza, Argentina \\
$^{3}$ Instituto de Ciencias Astron\'omicas, de la Tierra y del Espacio (ICATE), CC 49, CP 5400, San Juan, Argentina \\
$^{4}$ Instituto de  Astronom\'\i a Te\'orica y Experimental (IATE), Observatorio Astron\'omico, Universidad Nacional de C\'ordoba,\\ Laprida 854, X5000BGR, C\'ordoba, Argentina
}

\date{\today}

\begin{document}
\pagerange{\pageref{firstpage}--\pageref{lastpage}}

\maketitle

\label{firstpage}

\begin{abstract}

We analyse interacting galaxy pairs with evidence of tidal features in the Sloan Digital Sky Survey Data Release 7 (SDSS-DR7). 
The pairs were selected within $z<0.1$ by requiring a projected separation $r_p < 50 \kpc$  and relative radial velocity
$\Delta V < 500 \kms$. We complete spectroscopic pairs using galaxies with photometric redshifts considering $\Delta V_{phot} < 6800 \kms$, taking into account the mean photometric redshift uncertainty. 
We classify by visual inspection pairs of spirals into co-rotating and counter-rotating systems.  
For a subsample of non-AGN galaxies, counter-rotating pairs have larger star formation rates, and a higher fraction of young, star-forming galaxies. These effects are enhanced by restricting to $r_p < 12 \kpc$.
The distributions of $C$, $D_n(4000)$ and $(M_u-M_r)$ for AGN galaxies show that counter-rotating hosts have bluer colours and younger stellar population than the co-rotating galaxies although  the relative fractions of Seyfert, Liner, Composite and Ambiguous AGN  are similar. Also, counter-rotating hosts have more powerful AGN as revealed by enhanced $Lum[OIII]$  values. 
 The number of co-rotating systems is approximately twice the number of counter-rotating pairs which could be owed to a more rapid evolution of counter-rotating systems, besides possible different initial conditions of these interacting pairs.

\end{abstract}


\section{Introduction}

Galaxy interactions are one of the main mechanisms claimed for inducing star formation \citep{yee, kenni}, where starbursts are fuelled by gas inflows produced by the tidal torques generated during the encounters. \citet{barton}  and \citet{lam03} performed statistical analysis of star formation activity of galaxy pairs showing that proximity in radial velocity and projected distance are correlated with an increase in the star formation activity. \citet{alo06} analyse close pairs and found an excess of blue galaxies with respect to the control sample and associate this phenomenon with a larger fraction of star forming galaxies. They also found that in close pairs, there is an increment in the fraction of red galaxies compared to systems without a near companion. The authors propose a scenario where galaxies in pairs have formed stars efficiently  at early stages of their evolution, therefore, at present, they exhibit red colours. Another interpretation is that the presence of dust stirred up during encounters could affect colours and probably, partially obscure the star formation activity.  \citet{perez09} analysed the color distribution of close galaxy pairs finding an excess in both red and blue tails with respect to control samples. They also found that the trends still persists even after removing possible bias effects in the control sample selection. These results reinforce the claim that the excesses of blue and red galaxies are actually produced by galaxy interactions and not introduced by a biased selection.

\citet{alo12} performed an analysis of close galaxy pairs in groups and clusters, finding that these pairs tend to reside in groups with low density global environments. The authors also showed that pair galaxies have a significant excess of young stellar population with respect to group member galaxies without near companion. Recently, \citet{lam12} (hereafter L12) stressed the importance of studying different types of interactions and they classified pair galaxies into three categories: pairs undergoing merging $(M)$; pairs with evident tidal features $(T)$; and non disturbed pairs $(N)$. They also found an excess of galaxies in the blue peak in $M$ systems, while $T$ pairs show a larger fraction of galaxies in the red peak, compared to $N$ systems, specially for minor interactions.  These results suggest that the variation of the blue and red peak locations of the color bimodal distribution could be driven by different aspects of galaxy interactions such as evolutionary stage, gas content, interaction strength, etc. In particular L12 highlight the contribution of tidal interactions to the read peak in the galaxy pairs colour distribution. Therefore, it is interesting to focus in tidal interactions in order to analyse the role of this special type of encounters  on the main properties of galaxies in interacting systems.

Tidal interactions in galaxy pairs can produce several morphological features that are short lived such as extended structures,  bridges and tidal tails. Observations indicate that the probability to have grand-design arms is much higher for galaxies in binaries or groups than in the field \citep{kor79}. \citet{res} and references therein show  that the local fraction of tailed objects is about 1-2\% of the local galaxy population, and that the size of the tidal tails is related to the global dynamical structure of the interacting galaxies. \citet{moh}, points out that the tails in distant galaxies are shorter than those in nearby ones, reflecting the evolution with redshift of the sizes of spiral galaxies and their tidal structures. They also show that in simulated galaxies with low-mass halos, tidal tails turn out to be very long and the interacting galaxies merge very rapidly. Besides many authors have found objects such as dwarf galaxies and globular clusters in tidal debris from galaxy interactions \citep{mira, hun, char, smi}.

In this work we present an analysis of the galaxy properties in tidal interacting pairs, considering different types of morphological features that show this special type of systems. 
This paper is structured as follows: Section 2 describes the data used in this work. In Section 3 we show the procedure used to construct the tidal interacting pair catalogue, explaining the classification process of tidal pair galaxies, and we also present the general properties of the catalogue. An analysis of star formation rates, colours and stellar population, and their dependence on the different classifications in our sample of tidal interactions is described in Section 4. In addition, host active galaxy properties and nuclear activity are analysed in this section. Finally in Section 5, we summarize our main conclusions. 

Throughout this paper we adopt a cosmological model characterised by
the parameters $\Omega_m=0.3$, $\Omega_{\Lambda}=0.7$ and $H_0=100 \kms \rm Mpc ^{-1}$.

\section{Data}

Data Release 7 of Sloan Digital Sky Survey  (SDSS-DR7) \citep{dr7,sdss} is the seventh major data release, corresponding to the completion of the survey SDSS-II. It comprises $11.663$ sq. deg. of imaging data, with an increment of $\sim2000$ sq. deg., over the previous data release, mostly in regions of low Galactic latitude. SDSS-DR7 provides imaging data for 357 million distinct objects in five bands, \textit{ugriz}, as well as spectroscopy over $\simeq \pi$ steradians in the North Galactic cap and $250$ square degrees in the South Galactic cap. The average wavelengths corresponding to the five broad bands are $3551$, $4686$, $6165$, $7481$, and $8931$\AA\  \citep{fuku96,hogg01,smit02}. For details regarding the SDSS camera see \citet{gunn98}; for astrometric calibrations see \citet{pier03}.
The survey has spectroscopy over 9380 sq. deg.; the spectroscopy is now complete over a large contiguous area of the Northern Galactic Cap, closing the gap that was present in previous data releases.

In this work we analyse spectroscopic and photometric data extracted from SDSS-DR7. The spectroscopic data were derived from the Main Galaxy Sample (MGS; \citet{mgs}) obtained from the \texttt{fits} files at the SDSS home page\footnote{http://www.sdss.org/dr7/products/spectra/getspectra.html}. For this sample, k-corrections band-shifted to $z=0.1$, were calculated using the software \texttt{k-correct\_v4.2} of \citet{kcorrect}.
The photometric data were derived from the photometric catalogue constructed by \citet{photo}\footnote{http://casjobs.starlight.ufsc.br/casjobs/}. These authors compute photometric redshift and k-correction for the photometric data of the SDSS-DR7. The $rms$ of the photometric redshift is $\sigma_{phot} \sim$ 0.0227 and k-corrections were obtained through joint parametrisation of redshift and reference frame (at $z=0.1$) $(g-r)$ colour. For both data sets, k-corrected absolute magnitudes were calculated from Petrosian apparent magnitudes converted to the AB system.

\section{Interacting pair galaxies}

Galaxies in tidal pairs are subject to diverse perturbations that may alter their morphology and generate peculiar features such as bridges and tails. This diversity of structures can be produced by differences in the internal nature of the galaxies in the tidal pair interactions. For this reason, the aim of this section is to present a sample of tidal pairs in order to explore the different types of such interactions.

\begin{figure*} 
 \centerline{\psfig{file=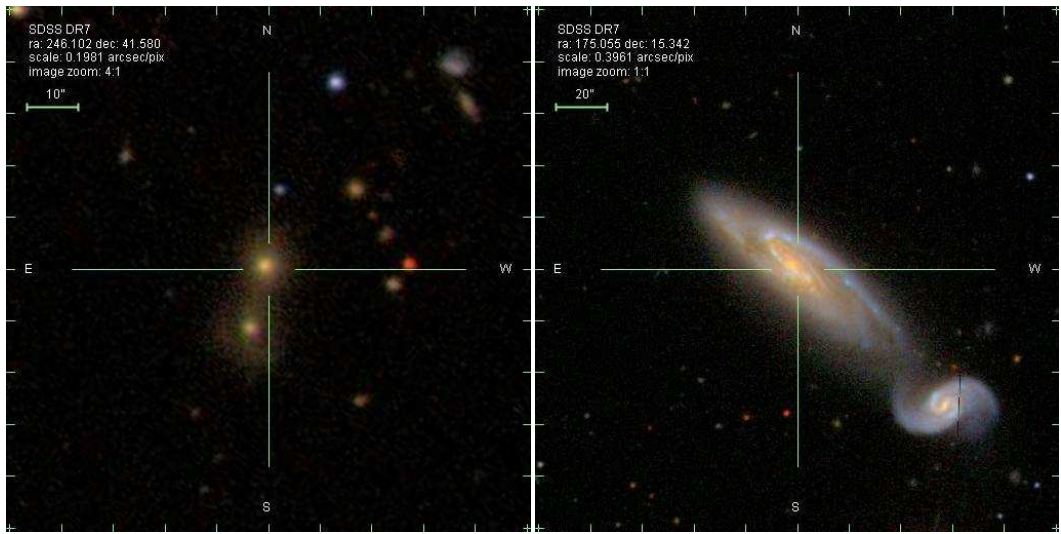,width=14cm,height=7cm}}
\caption{Examples of tidal galaxy pairs images. Left: galaxy pairs with a bridge (Tb). Right: interactions with tidal tails (Tt).  }
\label{ej1}
\end{figure*}

\subsection{Spectroscopic galaxy pair catalogue}

Some recent statistical work focus on the construction of galaxy pair catalogues with different projected separation ($r_{p}$) and relative radial velocity limits ($\Delta V$).
\citet{lam03} found that galaxies with a close companion within $r_{p}< 25 \kpc$ and $\Delta V < 100  \kms$ showed a higher star formation activity than isolated galaxies with similar redshift and luminosity distributions.
\citet{elli10}, in an study of the effects of environment on interactions, constructed a sample of pairs  with $r_{p}< 80 \kpc$ and $\Delta V < 500 \kms$ selected from SDSS-DR4, finding that pairs in high density
environments are characterised by wider separations and larger values of $\Delta V$.  
More recently, \citet{scu} presented a sample of pairs from SDSS-DR7 with 
 relative projected separation $r_{p}< 80 \kpc$ and relative radial velocities $\Delta V < 300 \kms$.
These pairs show star formation rate (SFR) enhancements of 30$\%$ out to at least $80 \kpc$.

In our previous work (L12) we build a catalogue of 1959 galaxy pairs with projected separation $r_{p}< 25 \kpc$ and relative radial velocities $\Delta V < 350 \kms$, within $z < 0.1$, finding 589 pairs that show tidal features.
 In order to obtain a larger number of pairs, and therefore improve the statistics, we extended this sample to $r_{p}< 50 \kpc$ and $\Delta V < 500 \kms$, and consider pairs with $z < 0.1$, where $z$ is the redshift of the brightest galaxy member of the system. In this way we obtain a sample of 1283 pair galaxies with tidal signatures. 

We performed an eye-ball classification using the SDSS-DR7 imaging available in 
CasJobs\footnote{http://cas.sdss.org/astrodr7/} in order to distinguish between different classes of tidal interactions. 
We classified the sample taking into account the presence of either large scale tidal tails (Tt) or a connecting bridge (Tb). Thus, Tt types are either pairs of two spiral galaxies or composed by a spiral and an elliptical galaxy while most Tb systems consist of a pair of elliptical galaxies connected by a bridge.  We found that 85\% of the pairs were classified into these subsamples. The remaining pairs that do not fulfil these two categories were excluded from the present analysis.
Fig. \ref{ej1} shows examples of Tt and Tb pairs.  
The visual inspection of the pairs was performed by all the authors dividing the total sample in subsamples of similar number of pairs. The reliability of the classification was addressed by a comparison of the classification of all authors in a common subsample. This comparison provides a classification uncertainty  which we estimate of approximately 3\%.

Under these considerations the resulting spectroscopic tidal galaxy pair catalogue comprises a total of 1082 pairs.

\subsection{Spectro-Photometric galaxy pair catalogue}

The SDSS spectrograph uses fibres manually connected to plates in the telescope's focal plane. 
These fibres are  mapped  through a mosaic algorithm \citep{blanton2003} that optimises the 
observation of large-scale structures. Two fibres can not be placed closer than 55''  
\citep{mgs}, so 
for two objects with the same priority (such as two MGS galaxies) and whose centres are closer than 55'', 
the algorithm selects at random the galaxy which will be observed spectroscopically. 
There are regions where the plates overlap (about $ 30\%$ of the mosaic regions), 
in which both objects may be observed. Due to fibre collision the spectroscopic sample is affected by incompleteness.
The magnitude limit of spectroscopic objects is $r=17.77$, but not all galaxies brighter than this 
limit were observed. This issue becomes more important when 
analysing compact objects.

Recently, \citet{OMill2012} employed photometric redshift information to quantify this effect on the detection of 
triplets of galaxies in the SDSS-DR7. These authors conclude that photometric redshift provide very useful 
information, allowing to asses fibre collision incompleteness and complete the sample of triple systems at low redshift.

In order to recover pair systems lost due the fibre collision effect, we have include galaxies with 
photometric redshifts that have r-band magnitude $r<17.77$. For each galaxy with spectroscopic measurements in the 
redshift range $z<0.1$,  we have searched a photometric companion that have a projected distance 
$r_{p}< 50 \kpc$ and a relative radial velocity $\Delta V_{phot} < 6800 \kms$, 
this last value correspond to $c\times \sigma_{phot}$ where 
$\sigma_{phot}$ is the mean photometric redshift error and $c$ is the speed of light. \citet{OMill2012} found that the 1$\sigma_{phot}$ interval provides a suitable compromise between high completeness and low contamination in the detection of triplets of galaxies. Including photometric information allows us to increment the 
number of galaxies in the sample under analysis and, therefore, improve the statistical significance of our results.

Similarly to the spectroscopic catalogue, we perform an eye ball classification of this sample, 
finding 2043 spectro-photometric galaxy pairs with tidal signs. Apart from the restriction on $\Delta V_{phot}$, the existence of tidal signatures reinforce that the galaxy pairs in the spectro-photometric sample are physical systems. About 78\% of the pairs in this sample  were classified as Tt or Tb, the remaining pairs that do not fulfil these two categories were excluded from the analysis. The resulting spectro-photometric tidal galaxy pair catalogue comprises a total of 1581 pairs. 

Table 1 provides the classification and number of pairs in the spectroscopic and spectro-photometric tidal pair samples.

\begin{table}
\center
\caption{Number of Tt and Tb pairs in the spectroscopic and spectro-photometric tidal pair samples}
\begin{tabular}{|c c c c| }
\hline
Samples & Spectroscopic & Spectro-photometric & Total\\
\hline
Tt      &  525          &  944                & 1469  \\
Tb      &  557          &  637                & 1194  \\
\hline
\hline
Total   & 1082          & 1581                & 2663 \\
\hline
\end{tabular}
\end{table}

\subsection{Characteristics of the pair samples}

In order to analyse different physical properties of galaxies in tidal pairs, we select galaxies with spectroscopic measurements and cross-correlate them with the derived galaxy properties from the MPA-JHU emission line analysis for the SDSS-R7\footnote{Avaible at http://www.mpa-garching.mpg.de/SDSS/DR7/}. From this catalogue we use the star formation rate normalized to the total mass in stars, $log(SFR/M_*)$, taken from \cite{bri04}.
We also use the spectral index $D_n(4000)$, as an indicator of the age of stellar populations. 
This spectral discontinuity occurring at 4000\AA \ \citep{kau02} arises by an
accumulation of a large number of spectral lines in a narrow region of the spectrum, an effect that is important in the spectra of old stars.
We have adopted \citet{bal99} definition of  $D_n(4000)$  as the ratio of the average flux densities in the narrow continuum bands (3850-3950 \AA \  and 4000-4100 \AA). 
We also use total stellar masses ($M_*$) based on fits to the photometry \citep{salim2007}.

\begin{figure} 

\centerline{\psfig{file=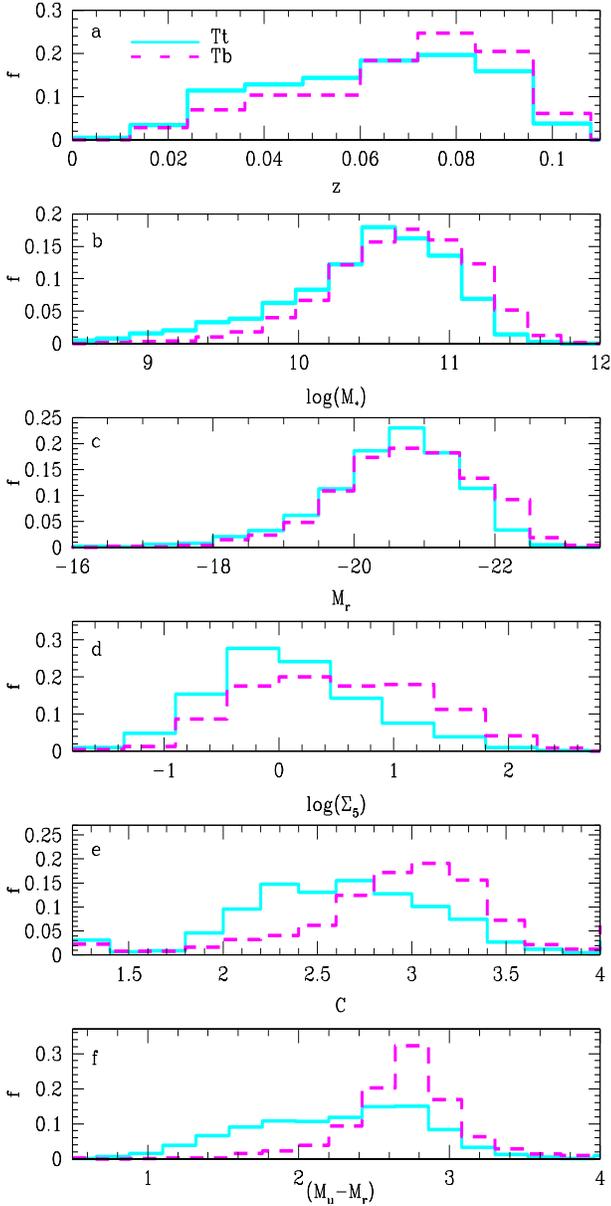,width=8cm,height=16cm}}
\caption{From top to bottom, ($a$) distribution of $z$, ($b$) $log(M_*)$, ($c$) $M_r$, ($d$) $log (\Sigma_5)$, ($e$) C  and ($f$) $(M_u-M_r)$ for galaxies in pairs classified as Tt (solid lines) and for Tb (dashed lines). }
\label{sig5}
\end{figure}

In Fig. \ref{sig5} we show different properties of Tt and Tb galaxy pair interactions (redshift, z, stellar mass content, $M_*$, absolute r-band magnitude, $M_r$, local density parameter, $\Sigma_5$, concentration index, $C$ and $(M_u-M_r)$ colour). 
From the upper panel it can be appreciated that both samples expand similar z ranges but the Tb pairs show a distribution slightly shifted towards higher z values. 
From panel $b$, we find that the Tb pairs show a stellar mass distribution 
 with small shift towards larger log($M_*$) values, than galaxies in the Tt sample. This behaviour is also reflected in the $M_r$ distribution (see $c$ panel).

With the purpose of analysing the local density environment of the galaxy pairs in our samples, 
 we have calculated the local density parameter $\Sigma_5$,
 to give an estimate of the mean environment of the two subsamples.  
This parameter is defined through the projected distance $d$ 
to the $5^{th}$ nearest neighbour brighter than $M_r < -20.5$ \citep{balo04}, $\Sigma_5 = 5/(\pi d^2)$,
with a radial velocity difference less than 1000$\kms$, and provides a suitable measurement of the local density of the systems. \citet{alo06} identified three regions according to the value of log$(\Sigma_5)$: low density (log$(\Sigma_5)<$ -0.57 ), medium density (-0.57 $<$ log$(\Sigma_5)<$ 0.05) and hight density (log$(\Sigma_5)>$ 0.05). 
The results displayed in Fig. \ref{sig5} ($d$ panel) show that Tb pairs present a higher fraction of galaxies residing in high density environments, while galaxies in Tt pairs are preferentially hosted in environments with intermediate densities.

The concentration index C\footnote{$C=r90/r50$ is the ratio of Petrosian 90 \%- 50\% 
r-band light radii} is a good morphological classification parameter \citep{C} and correlate with the stellar mass ($M_*$) and SFR \citep{c1}. \citet{c2} performed a galaxy morphological classification using the C parameter, finding a very good agreement with the visual classification. 
We adopted the critical concentration index value of $C=2.5$ to segregate concentrated, bulge-like galaxies ($C >2.5$) and more extended, disk-like objects ($C<2.5$).
C index distributions are showed in panel $e$. From this figure it can be appreciated that the sample of Tb pairs exhibiting higher values of C index with respect to the Tt sample. Moreover, we find that Tb pairs show a higher fraction of galaxies with $C>2.5$ ($\approx$ 85\%), with respect to the Tt sample 
($\approx$ 60\%). This result indicates that a higher fraction of galaxies in the Tb sample presents bulge morphology, while Tt pairs show a significant fraction of disk type objects. This result is expected under the classification scheme used in this work and because  elliptical galaxies are more compact and tend to inhabit higher density regions, so under these conditions is more difficult to develop tidal tails.

Nevertheless, the most significant difference between galaxies in Tb and Tt pairs resides in the $(M_u-M_r)$ colour distribution. About 95\%  of the Tb galaxies present red colours ($(M_u-M_r)>2.0$) while galaxies in Tt pairs present a more uniform distribution, with 68\% of galaxies with redder colours. This result is consistent with the previously found in the distribution of the concentration index, indicating that the sample of Tb pairs is composed mostly by redder early type galaxies.

The trends found with this analysis persist even when considering control samples matched in $z$, $M_r$, $M_*$ and $\Sigma_5$. This result suggests that the so called `red peak' present in the color distribution of pair galaxies, reported by different authors
\citep[e.g.][]{alo06,alo12,perez09,Darg10,patton11}, is mostly populated by this special type of tidal interaction that comprises galaxies with a bridge (Tb).  We will analyse this topic in detail in a forthcoming paper and, in what follows, we will consider only interactions with tidal tails (Tt) and in particular we will study the interactions between spiral galaxies within this subsample.


\section{Co-rotating and counter-rotating systems}

\begin{figure*}
 \centerline{\psfig{file=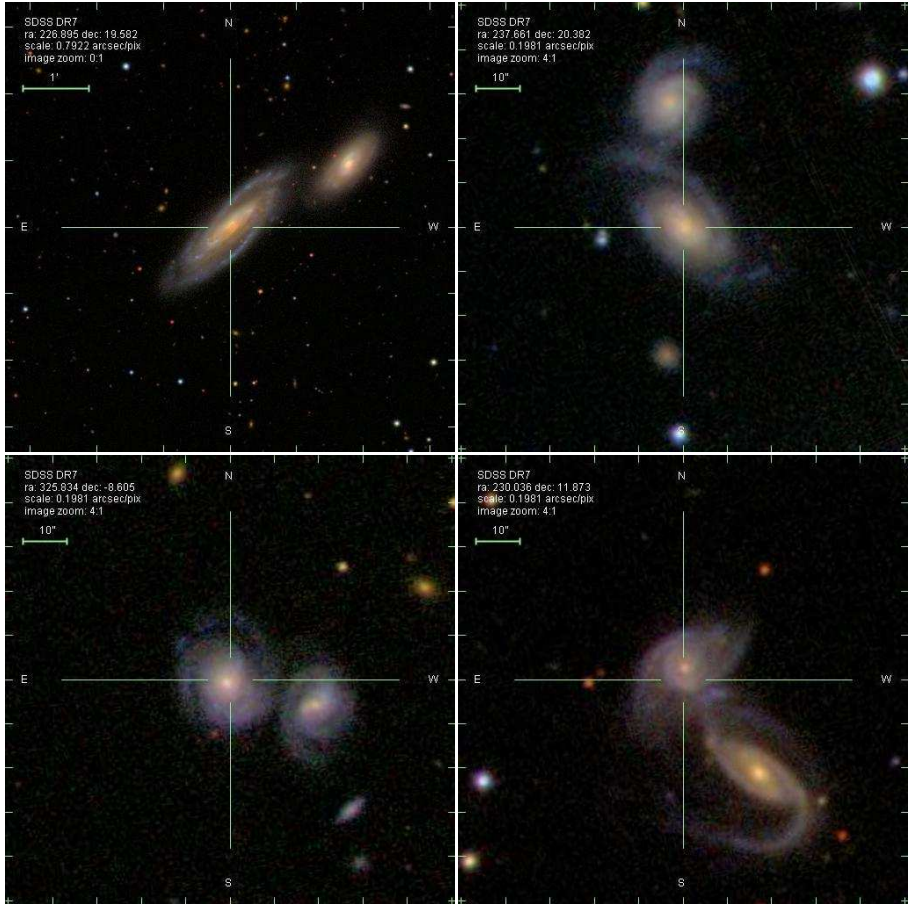,width=12cm,height=12cm}}
\caption{Examples of galaxy pair images with different classification: 
co-rotating (left) and counter-rotating (right) in the spectroscopic sample (upper panels) and spectro-photometric catalogue (bottom panels). }
\label{ej}
\end{figure*}

Numerical simulations show that prograde encounters between equal mass galaxies represent the most favourable scenario for creating tidal tails \citep{toom, dub}, instead,  retrograde encounters have greater star formation efficiency than direct encounters.  \citet{dimm07} analysed  direct and retrograde galaxy encounters finding that galaxies with opposite angular momentum (retrograde encounters) are less affected by tidal interactions and the gas mass content remain confined in the disk forming a reservoir for the intense starburst in the final stage of the merging.  
On the other hand, \citet{cer} found a correlation between the spin magnitude of neighbouring galaxies, but no clear alignment between orientation. Regarding halo masses, \citet{dub} found that it is difficult to form a long tail in collisions of galaxies with very massive halos.

In the following study we only analyse tidal interactions between spiral galaxies with tidal tails within Tt pairs, and we visually sub-classified these pairs according to the sense of rotation of the spiral arms. Therefore, we have defined two categories: co-rotating and counter-rotating pairs. If both  galaxies have the  same spiral pattern sense we call the system co-rotating, and otherwise we call the system counter-rotating. Analogously to section 3.1 we estimate the uncertainty in this classification that results in 4\%. It is noteworthy that the visual inspection performed in order to identify the sense of rotation of the galaxy spiral arms, produces a selection effect and, therefore, most of the systems under study in this section are major mergers. We have analysed how these different spin configurations affect galaxy luminosities, star formation rates, spectral indicators of stellar populations, colours and active nuclei properties. 

 Upper panels of Fig. \ref{ej} show images of typical examples of co-rotating and  counter-rotating pairs in the spectroscopic sample. In a similar way, bottom panels show images of typical examples of these categories in the spectro-photometric catalogue. 
 
Table 2 provides the percentages of co-rotating and counter-rotating pairs 
in the spectroscopic and spectro-photometric samples. Remarkably, we find that co-rotating systems doubles the number of the counter-rotating pairs. A possible explanation for this fact could reside in the initial conditions of nearby halos and that counter-rotating interactions are likely to be more violent, with mergers occurring faster, and with less prominent large-scale tails \citep{her04}. We notice that this effect is present in both samples, spectroscopic and spectro-photometric catalogues, indicating that the results are not biased by the inclusion of photometric galaxies.

\begin{table}
\center
\caption{Classification, number of pairs and percentages of co-rotating and counter-rotating galaxy pairs in the spectroscopic and spectro-photometric samples (top and bottom, respectively)}
\begin{tabular}{|c c c c| }
\hline 
Sample & Classification & Number of pairs &  Percentages  \\
\hline 
\hline
Spectroscopic & Co-rotating          &   171       &  65.4\%         \\
              & Counter-rotating      &   85 &     34.3\%     \\
\hline 
                         & Total             &       256    &  100\%   \\
\hline
\hline\
Spectro- 			& Co-rotating          &    266      &   66.8\%      \\
photometric                    & Counter-rotating    &   139  &   33.2\%       \\
\hline 
                       & Total             &      405     &  100\%   \\
\hline 
\end{tabular}
{\small  }
\end{table}

\subsection{Galaxies without nuclear activity and AGN hosts }

In the following analysis, we used galaxies with spectroscopic measurements and distinguish between active galactic nuclei (AGN) and galaxies without a detected AGN (hereafter non-AGN galaxies) in co and counter-rotating pairs. 
For the AGN selection we use the publicly available SDSS emission-line fluxes. The method for emission-line measurement is detailed in \cite{tremonti04}.  Additionally, we have corrected the emission-line fluxes for optical reddening using the Balmer decrement and the \cite{calzetti00} dust curve.  We assume an $R_V=A_V/E(B-V)=3.1$ and an intrinsic Balmer decrement $(H\alpha/H\beta)_{0}=3.1$ \citep{OM89}.  
Since the true uncertainties in the emission-line measurements were underestimated the signal-to-noise ($S/N$) of every line was calculated with the emission-line flux errors adjusted according to the uncertainties suggested by the MPA/JHU catalogue.

The AGN galaxy sample was selected using a standard diagnostic diagram proposed by \cite{BPT81} 
(hereafter BPT). This diagram allows the separation of type 2 AGN, from normal star-forming 
galaxies, using emission-line ratios and depending on their position in the diagram.
Furthermore, we used only galaxies with signal-to-noise ratio S/N $> 2$ for all the lines
intervening in the diagnostic diagram used to discriminate AGN from HII galaxies.
This S/N cut was selected taking into account that the adjusted uncertainties almost duplicated
the original errors.
So, taking into account the relation between spectral lines, $\rm [OIII]\lambda 5007$, $\rm H\beta$,
$\rm [NII]\lambda 6583$ and $\rm H\alpha$, within the BPT diagram we follow the \cite{kauff03} 
criterion to select type 2 AGN as those with:   

\small
\begin{equation}
\log_{10}([OIII]/\rm H\beta) > 0.61/(\log_{10}(\rm [NII/H\alpha])-0.05)+1.3
\end{equation}
\normalsize \

using this criteria we build two catalogues, a sample of galaxies free of AGN and 
a sample of AGN galaxies. Under these restrictions, and from a total of 917 interacting galaxies with these 
spectroscopic measurements, we found 297 AGN, representing a fraction of about 30$\%$. Table 3 summarizes the 
percentages. It is important to highlight that the percentages of co and counter-rotating systems in the sample of non-AGN galaxies and in the sample of AGN are similar.

\begin{figure}
\begin{picture}(200,227)
\put(0,0){\psfig{file=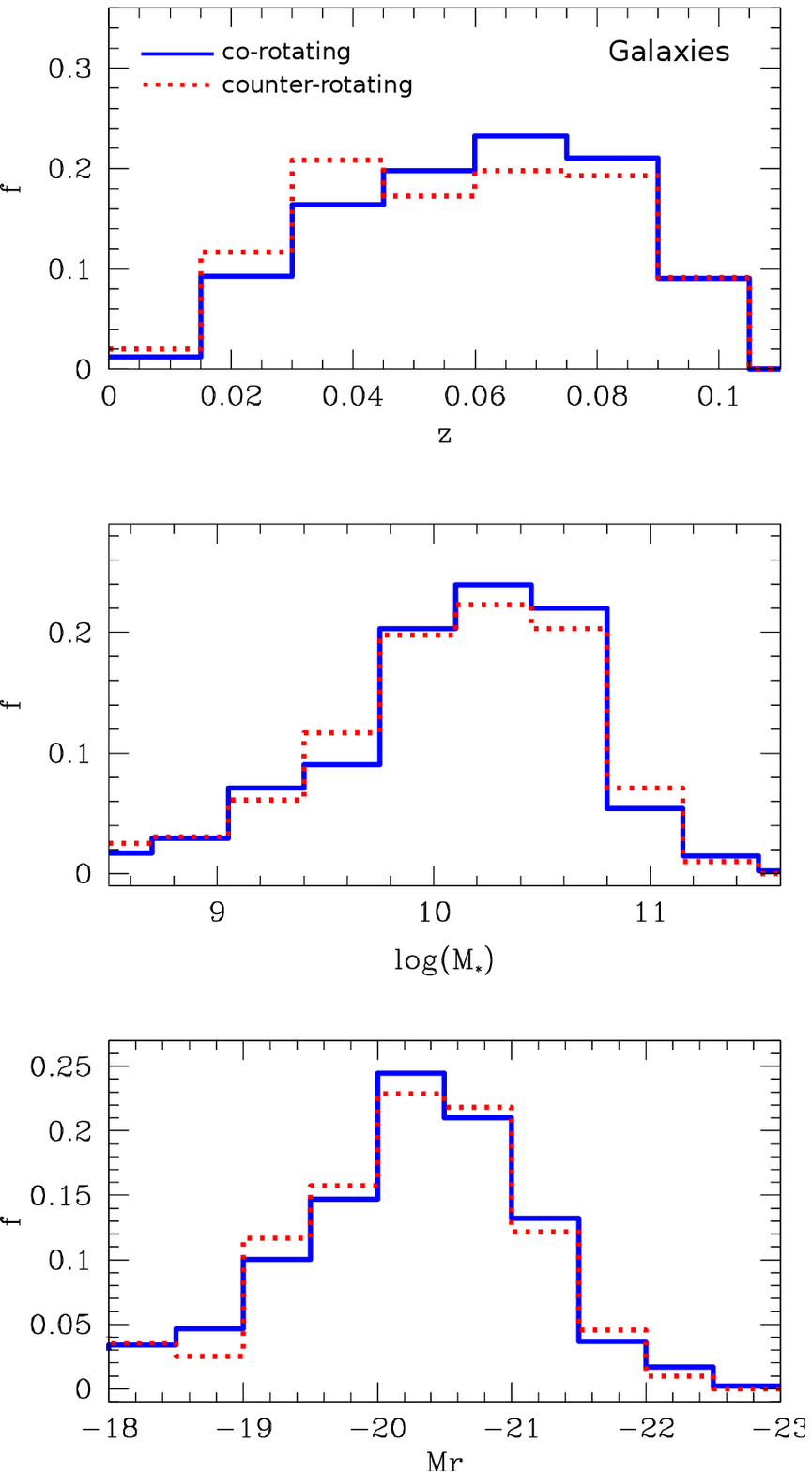 ,width=4cm,height=8cm}}
\put(125,0){\psfig{file=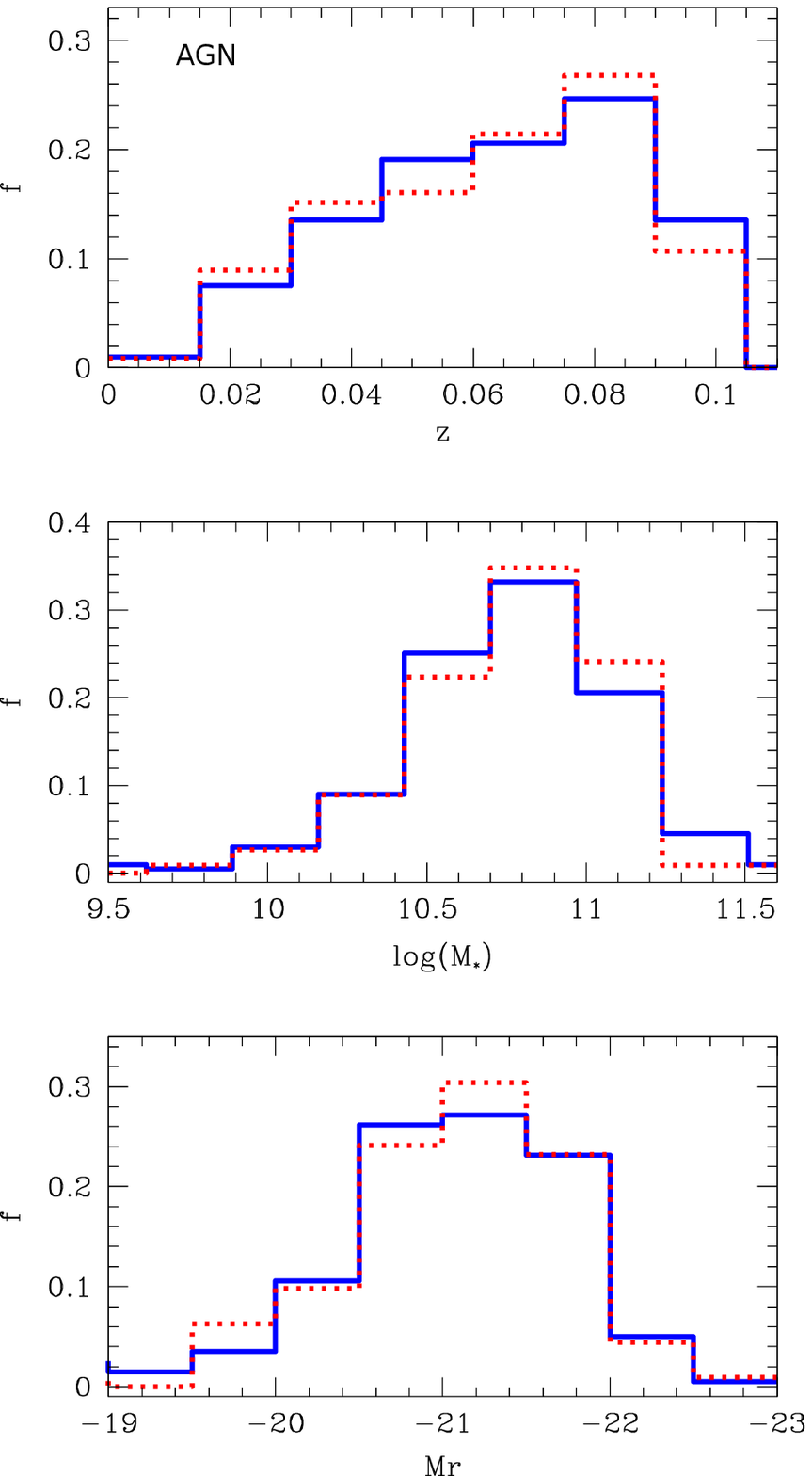, width=4cm,height=8cm}}
\end{picture}
  \caption{Distribution of redshift (top), absolute r-band magnitude (middle) and stellar mass content (bottom) for the non-AGN galaxy sample (left) and for the AGN galaxy sample (right) in co and counter-rotating pairs. 
The solid lines correspond to the co-rotating and the dotted lines correspond to the counter rotating galaxy pairs, respectively.}
  \label{f4}
  \end{figure}

In Fig. \ref{f4} we show the distribution of redshift, stellar mass content and r-band absolute magnitude 
for the sample of non-AGN galaxies (left) and for the sample of AGN galaxies (right) in co and counter-rotating pairs (solid and dotted lines, respectively). 
From this figure it can be appreciated that both samples present similar  z, $M_*$ and $M_r$ distributions.

\begin{table}
\center
\caption{Classification, number of galaxies and percentages of co-rotating and counter-rotating galaxy pairs in the AGN sample (top), and in the galaxy sample (bottom)}
\begin{tabular}{|c c c c| }
\hline 
Sample & Classification & Number of galaxies &  Percentages  \\
\hline 
\hline
AGN & Co-rotating          &    191      &  64.3\%         \\
              & Counter-rotating      &  106  &    35.7\%     \\
\hline 
                         & Total             &    297       &  100\%   \\
\hline
\hline\
non-AGN galaxies			& Co-rotating          &    417      &  67.3\%      \\
                    & Counter-rotating    &  203   &  32.7\%       \\
\hline 
                       & Total             &      620     &  100\%   \\
\hline 
\end{tabular}
{\small  }
\end{table}

\subsubsection{Galaxy properties}

In this section we focus our analysis on the sample of spiral galaxies with tidal tails in order to investigate the differences in the properties of the systems according to the sense of rotation of the member galaxy pairs. We study the sample of non-AGN galaxies, defined in the previous subsection, and analyse separately co-rotating and counter-rotating interactions and their effects on the galaxy properties. We also analyse a subsample of closer encounters considering pairs with projected separations $r_p < 12 \kpc$. This value corresponds to the the median of the $r_p$ distribution for the total sample. 

We analyse the specific star formation rate, $SFR/M_*$, the $D_n(4000)$ parameter, the $(M_u-M_r)$ colour index and the concentration parameter C.  

\begin{figure*}
\begin{picture}(350,450)
\put(10,10){\psfig{file=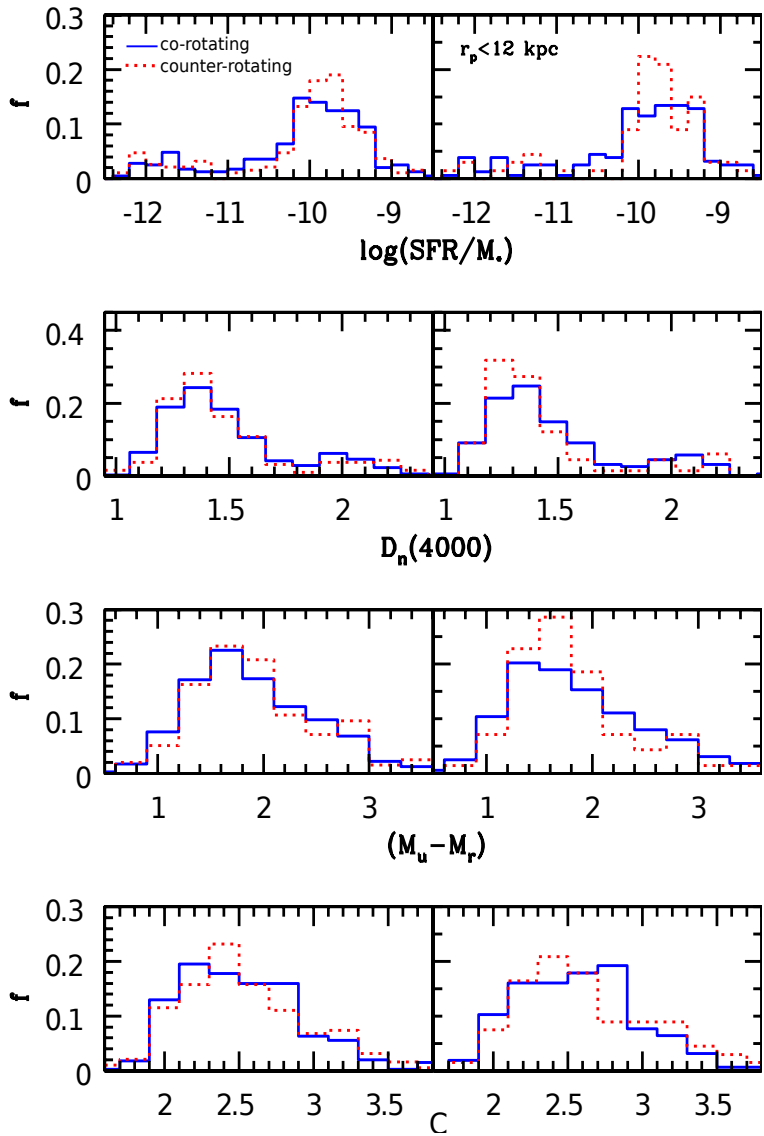,width=10cm,height=15cm}}
\end{picture}
\caption{Distribution of $log(SFR/M_*)$, $D_n(4000)$, $(M_u-M_r)$ and $C$  
for pair galaxies, excluding AGN, classified as co and counter-rotating  
(solid and dotted lines, respectively).  
}
\label{hist}
\end{figure*}

The distribution of these parameters are shown in Fig. \ref{hist} for the two samples with $r_p< 50 \kpc$ (left) and $r_p < 12 \kpc$ (right). From this figure it can be appreciated that the counter-rotating systems show a higher star formation rate. This effect is more strongly seen in the sample with $r_p < 12 \kpc$, where the effect is also reflected in the lower values of $D_n(4000)$ parameter and $(M_u-M_r)$ colours. Both samples present similar C values.

\begin{table*}
\center
\caption{Percentages of galaxies with $log(SFR/M_*)>-10.0$, $D_n(4000)<1.4$, $(M_u-M_r)<2.0$ and $C<2.5$ in co-rotating and counter-rotating  systems. }
\begin{tabular}{c c c c c}
\hline
& \multicolumn{2}{c}{Pairs with $r_p< 50 \kpc$} &   \multicolumn{2}{c}{Pairs with $r_p< 12 \kpc$}   \\
\hline
  Ranges & Co-rotating $\%$ &  Counter-rotating $\%$ & Co-rotating $\%$ &  Counter-rotating $\%$    \\       
\hline
\hline
 $log(SFR/M_*)>-10.0$ & 52.7 $\pm$ 2.5  &  59.0 $\pm$ 3.4  & 57.3 $\pm$ 3.8  &  72.2 $\pm$ 5.5  \\
\hline
  $D_n(4000) < 1.4$          &  49.92 $\pm$ 2.6    &  55.4  $\pm$ 3.5 &  54.7 $\pm$ 3.9    &  68.0  $\pm$ 5.5  \\
\hline
  $(M_u-M_r)<2.0$           & 62.4 $\pm$ 2.4  &  62.0 $\pm$ 3.4  & 65.4 $\pm$ 3.7  &  72.4 $\pm$ 5.4   \\
\hline 
  $C<2.5$       & 52.2  $\pm$ 2.5  & 54.0 $\pm$ 3.6  & 46.2  $\pm$ 3.8  & 49.3 $\pm$ 6.1  \\
\hline
\end{tabular}
{\small  }
\end{table*}

In order to quantify these differences we have estimated the fraction of galaxies with $log(SFR/M_*)>-10.0$, $D_n(4000)<1.4$ and $(M_u-M_r)<2.0$ and $C<2.5$. In Table 4 we summarize these results. All the uncertainties in this paper were derived through a bootstrap resampling technique \citep{efron}. It can be seen a larger star formation rate with an associated younger stellar population in counter-rotating systems, an effect that is stronger for the closest $r_p < 12 \kpc$ subsample, with a significance of $\sim 2.5\sigma$. We argue that this larger star formation activity in counter-rotating systems can be related to the larger disturbing effects of such interactions in comparison to those in co-rotating pairs. According to simulations of disk-type galaxy interactions, there is a tight correlation of the starburst and the pericentre passage \citep{Mihos94,dimm07}. Statistically, low $r_p$ values could have a larger chance of being galaxies at the pericentre distance, giving rise to the stronger effects for the $r_p < 12 \kpc$ subsample.

\subsubsection{AGN and host properties}

In this section we use the sample of AGN described in section 4.1, with the purpose of 
analysing the possible differences in the properties of AGN hosts in co and counter-rotating tidal galaxy pairs. We also investigate the influence of the sense of rotation of tidal interacting spiral galaxies on the black hole activity.

\begin{figure}
\centerline{\psfig{file=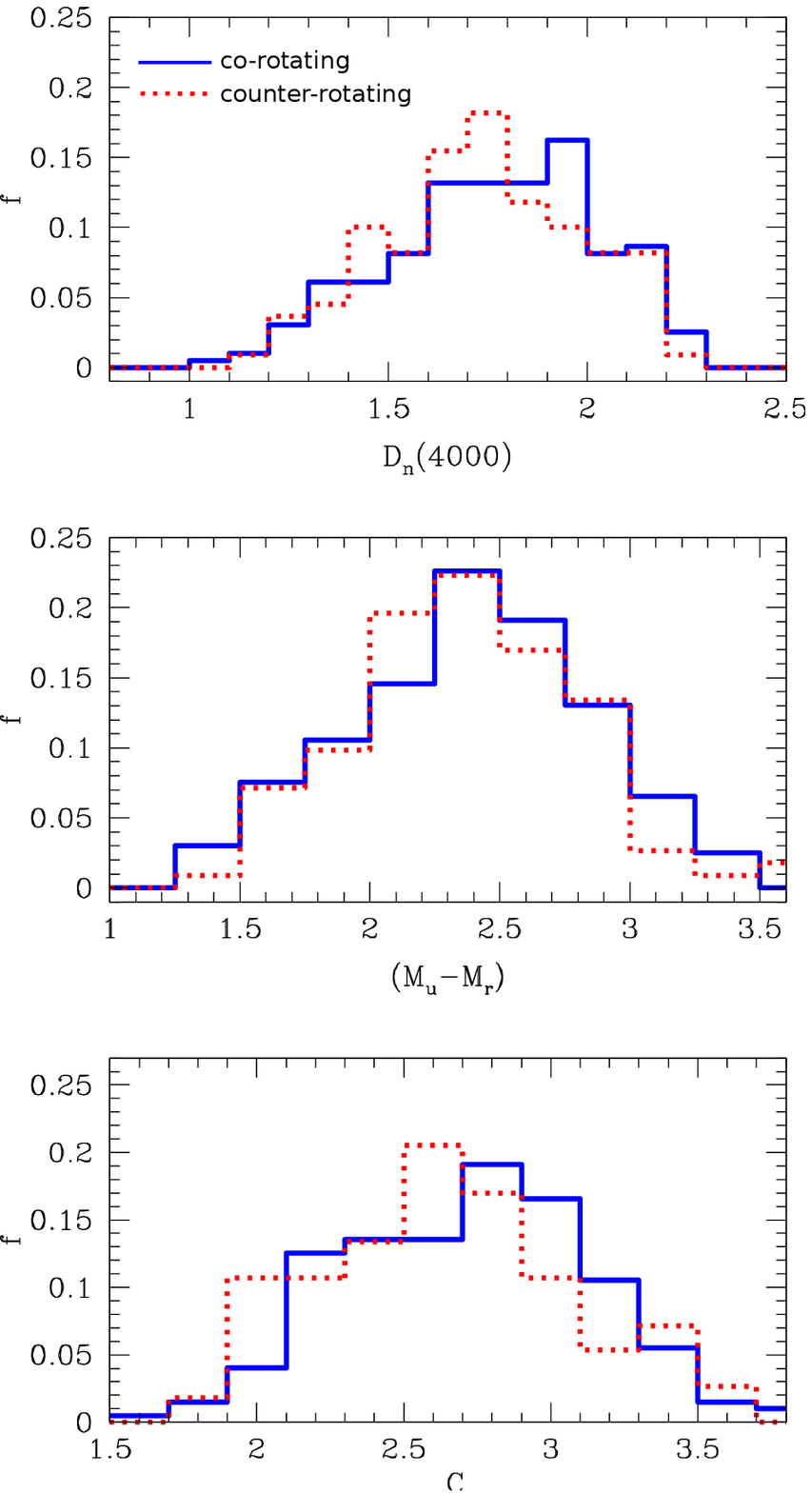,width=7cm,height=12cm}}
\caption{Distribution of $C$, $D_n(4000)$ and ($(M_u-M_r)$) for AGN pair galaxies classified as co and counter-rotating (solid and dotted lines, respectively).  
}
\label{hist1}
\end{figure}

In Fig. \ref{hist1} we show the distribution of $C$, $D_n(4000)$ and $(M_u-M_r)$ colour for co and counter-rotating AGN host galaxies. This figure shows only a slightly difference in the behaviour of AGN hosts for the two subsamples, with counter-rotating galaxies presenting bluer colours and younger stellar population than their co-rotating counterpart. In order to quantify these difference, we estimate the fraction of galaxies with $D_n(4000)<1.8$, $(M_u-M_r)<2.3$ and $C<2.7$. 
The thresholds  were chosen taking into account that, on average, AGN pairs with tidal features have larger 
$D_n(4000)$ values and redder colours \citep{agn1}. In addition, \citet{c1} showed that AGN are preferentially found in more concentrated host galaxies. 
The results are shown in Table 5, and it can be appreciated similar tendencies that in the galaxy sample analysed in the previous section.

\begin{table}
\center
\caption{Percentages of AGN galaxies with $D_n(4000)<1.8$, $(M_u-M_r)<2.3$ and $C< 2.7$ in 
co-rotating and counter-rotating  hosts. }
\begin{tabular}{|c c c | }
\hline
Ranges & Co-rotating $\%$ &  Counter-rotating $\%$    \\
\hline
\hline
$D_n(4000) < 1.8$     &  51.7 $\pm$ 3.5    &  61.1  $\pm$ 4.8   \\
\hline
$(M_u-M_r)<2.3$           &  38.6 $\pm$ 3.4    &  45.9  $\pm$ 4.7   \\
\hline 
$C<2.7$               & 45.3  $\pm$ 3.5    &  57.0  $\pm$ 4.5   \\
\hline
\end{tabular}
{\small  }
\end{table}

\begin{figure} 
\centerline{\psfig{file=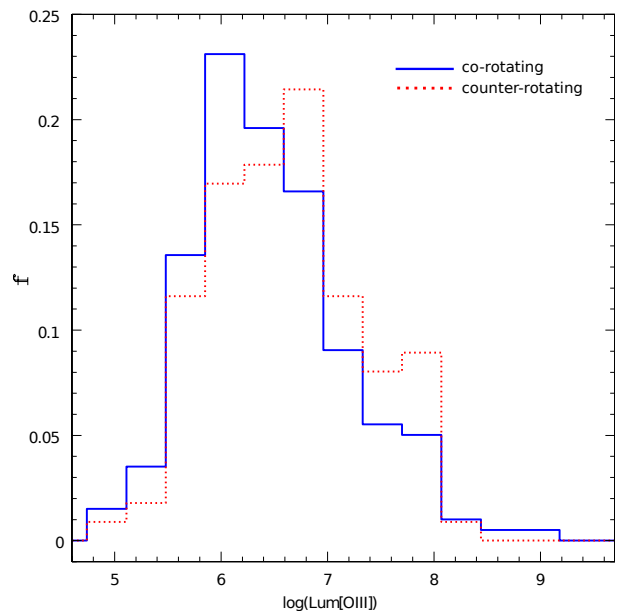,width=8cm,height=8cm}}
\caption{ Distribution of log($Lum[OIII]$) for AGN galaxies classified as
co-rotating (solid lines) and counter-rotating (dotted lines). 
}
\label{agn}
\end{figure}

As a tracer of the AGN activity, we focus here on the dust corrected luminosity of 
the [OIII]$\lambda$5007 line, $Lum[OIII]$.
This estimator is widely used by several authors  \citep{mul94, kauff03, heck04, heck05, 
bri04} mainly due that the [OIII] line is one of the strongest narrow emission 
lines in optically obscured AGN and with very low contamination by contributions of 
star formation in the host galaxy. 

The distributions of the AGN power derived from the OIII luminosity of the two subsamples are shown in Fig. \ref {agn}, and it can be appreciated that counter-rotating AGN galaxies present a slight excess of high log($Lum[OIII]$) values, with respect to the co-rotating counterpart.
In order to quantify these excesses, we calculate the fraction of galaxies with $L[OIII]>6.5$ for both samples, 
finding $21.03 \pm 2.92 \%$ and $27.71 \pm 4.18 \%$ for co and counter-rotating galaxies,
respectively. This result could be suggesting that AGN host galaxies in counter-rotating pairs are more powerful 
than those in the co-rotating systems.

\begin{figure*}
\centerline{\psfig{file=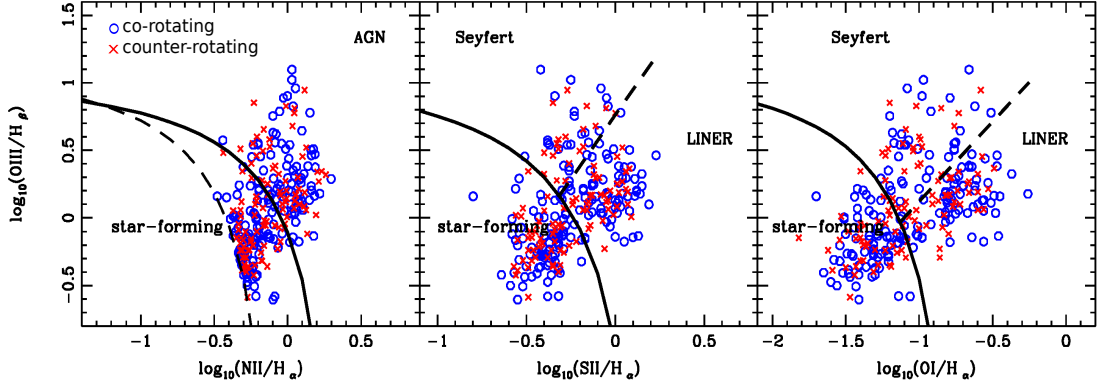,width=16cm,height=6cm}}
\caption{The three BPT diagrams used to classify the emission-line galaxies as: Seyfert, LINER, Composite and 
Ambiguous galaxies.
Left: Dashed line shows the Kauffmann et al (2003) selection criteria. The Kewley 
et al. (2006)  classification is shown as the solid line.
Middle and Right: The solid lines separate the star-forming galaxies from the active galaxies and the 
dashed lines represent the Seyfert-LINER demarcation.}
\label{bpt}
\end{figure*}

Moreover, the AGN sample can be classified into Composite, 
Seyfert, LINER and Ambiguous following the classification procedure proposed by \cite{Kewley01,Kewley06}. 
In this scheme \cite{Kewley01} define a theoretical discrimination between starburst regions from objects of other types of excitation using three BPT diagrams, as shown in Fig. \ref{bpt}  (solid lines). Thus, this AGN classification depends on the relative location  following equations 2 to 4:  

\small
\begin{equation}
\log_{10}([OIII]/\rm H\beta) > 0.61/(\log_{10}(\rm [NII/H\alpha])-0.47)+1.19
\end{equation}
\begin{equation}
\log_{10}([OIII]/\rm H\beta) > 0.72/(\log_{10}(\rm [SII/H\alpha])-0.32)+1.30
\end{equation}
\begin{equation}
\log_{10}([OIII]/\rm H\beta) > 0.73/(\log_{10}(\rm [OI/H\alpha])+0.59)+1.33
\end{equation}
\normalsize \

In addition, \cite{Kewley06} provide an empirical division between Seyfert
and LINER galaxies given by the following parametrization, 
where {\bf Seyfert} galaxies are located above the Seyfert-LINER line (dashed line in Fig. \ref{bpt}b and Fig \ref{bpt}c) on the $SII/H\alpha$ and $OI/H\alpha$ diagrams:

\small
\begin{equation}
\log_{10}([OIII]/\rm H\beta) > 1.89\log_{10}(\rm [SII/H\alpha])+0.76
\end{equation}
\begin{equation}
\log_{10}([OIII]/\rm H\beta) > 1.18\log_{10}(\rm [OI/H\alpha]+1.30
\end{equation}
\normalsize \

and {\bf LINER} galaxies are located below the Seyfert-LINER line:

\small
\begin{equation}
\log_{10}([OIII]/\rm H\beta) < 1.89\log_{10}(\rm [SII/H\alpha])+0.76
\end{equation}
\begin{equation}
\log_{10}([OIII]/\rm H\beta) < 1.18\log_{10}(\rm [OI/H\alpha]+1.30
\end{equation}
\normalsize \

Also, galaxies lying in between the \cite{kauff03} and \cite{Kewley01} classification lines are 
classified as {\bf Composite} galaxies. These galaxies seem to be in a transition between the HII-region and AGN, 
and probably contain a mixture of metal-rich stellar population plus AGN emission. 
Finally, {\bf Ambiguous} galaxies are those classified as one type of object in one BPT and another in the 
remaining two diagrams. 
In Fig. \ref{bpt} the relative position of the AGN in co and counter rotating pairs are shown in the 
three BPT diagnostic diagrams.
As it can be appreciated, there is not evidence of differences
in the position of AGN in co and counter rotating pairs regarding this classification scheme.

The results are quantified in columns $2$ and $4$ of Table 6 where the percentage of co and counter rotating 
galaxy pairs is calculated for every type of object.

\begin{table*}
\center
\caption{Percentages of co-rotating and counter-rotating AGN galaxies in tidal pairs and median values of ($Lum[OIII]$) for every type of object classification.}
\begin{tabular}{|c c c c c | }
\hline
  & \multicolumn{2}{c}{Co-rotating} & \multicolumn{2}{c}{Counter-rotating}   \\
\hline
 Classification & $\%$ & $<Lum[OIII]>$ & $\%$ & $<Lum[OIII]>$ \\
\hline
\hline
Seyfert    &  12.9 $\pm$ 2.9 & 7.1 $\pm$ 0.4 &  14.1 $\pm$ 3.5 & 6.9 $\pm$ 0.5 \\
Liners     &  18.0 $\pm$ 3.9 & 6.1 $\pm$ 0.3 &  19.8 $\pm$ 4.4 & 6.2 $\pm$ 0.3\\                
Composite  &  54.1 $\pm$ 7.3 & 6.4 $\pm$ 0.2 &  50.0 $\pm$ 6.9 & 6.6 $\pm$ 0.2\\           
Ambiguous  &  15.0 $\pm$ 3.5 & 6.2 $\pm$ 0.2 &  16.1 $\pm$ 3.8 & 6.6 $\pm$ 0.3\\           

\hline
\end{tabular}
{\small  }
\end{table*}

In addition, we have explored which of the AGN subclasses present higher median values of $Lum[OIII]$. The results are shown in columns $3$ and $5$ of Table 6. It can be noticed that galaxies classified as Composite and Ambiguous present noticeable differences between co and counter-rotating galaxy pairs for median values of $Lum[OIII]$ being  higher for the counter-rotating ones. The median of $Lum[OIII]$ is almost identical (within the estimated errors) for Seyfert and LINER galaxies. This result suggests that these galaxies which have a dubious AGN classification are those which have the strongest differences depending of the sense of rotation. This fact favours the interpretation of the different $Lum[OIII]$ values according to the sense of rotation, as differences in the star formation activity rather than the AGN itself. 
In this sense, \cite{heck04}, compute the average contribution of the AGN to the $Lum[OIII]$ for a sample of Composite galaxies and the AGN-dominated galaxies taken from SDSS. The results of their work indicate that the $Lum[OIII]$ emission from Composite galaxies comes, in a range of 50\% to 90\% from AGN activity, while for AGN-dominated galaxies, more than 90\% of the $Lum[OIII]$ comes from AGN emission.  Once nuclear activity is triggered, the co-rotating vs. counter-rotating large scale structure becomes irrelevant, nuclear activity is a controlled by very localized small  scale physics, see for instance \citet{cer11}.

\section{Conclusions}

We have performed a statistical analysis of interacting pairs with tidal features, selected from SDSS-DR7. We obtained a spectroscopic sample considering $r_p < 50 \kpc$ and 
$\Delta V < 500 \kpc$. With the aim to obtain better statistical results, for each galaxy with spectroscopic measurements  we have searched a photometric companion with projected distance 
$r_{p}< 50 \kpc$ and a radial velocity difference $\Delta V_{phot} < 6810 \kms$. First, we divided the sample taking into account the presence of tidal tails (Tt) or a connecting bridge (Tb). These subsamples exhibit similar distributions of $z$, $M_r$ and $log(M_*)$, while local environment, concentration index and color distributions are very different, showing that Tb pairs have a higher fraction of galaxies residing in high density environments, present redder colours and exhibit a bulge-type morphology. 
The results found in this analysis persist even when considering control samples matched in $z$, $M_r$, $M_*$ and $\Sigma_5$, suggesting that the so called `red peak' present in the color distribution of pair galaxies, is mostly populated by tidal interaction that comprises galaxies with a bridge (Tb). 
In this work, we analyse in detail interactions with tidal tails (Tt) between spiral galaxies, and we sub-classified these pairs according the sense of rotation of the spiral arms, by visual inspection of the images. Then, we defined two categories: co-rotating and counter-rotating pairs. 

We can summarize the main results in the following conclusions.

\begin{itemize}

\item We find that co-rotating systems doubles the number of the counter-rotating pairs, an effect that is present in both samples, spectroscopic and spectro-photometric catalogues.
This could be due a combination of parallel spin preference in initial conditions of nearby halos, a more rapid evolution to merger, and a less effective tidal tail development.

\item We have studied separately galaxies without nuclear activity, and AGN hosts in co and counter-rotating tidal pairs. The percentages of co and counter-rotating tidal systems show the same result as mentioned above in both pair samples. We also find similar distributions of redshift, stellar mass content and r-band absolute magnitude for samples of AGN and non-AGN galaxies in co and counter-rotating pairs.

\item For galaxies with no AGN activity we find that counter-rotating systems have a higher star formation rate and a  younger stellar population. This effect is stronger in the sample with $r_p < 12 \kpc$, as expected due to the larger tidal effects associated to closer encounters.
 
\item We find that the distributions of $C$, $D_n(4000)$ and $(M_u-M_r)$ of AGN host galaxies present a small difference in the behaviour of the two subsamples, showing that counter-rotating hosts have a higher fraction of galaxies with bluer colours and younger stellar population than their co-rotating counterpart. 

\item The AGN sample was classified into Composite, Seyfert, LINER and Ambiguous following the \cite{Kewley01,Kewley06} scheme finding no significant  differences in the position of AGN in co and counter rotating pairs regarding this classification.

\item As a tracer of the AGN activity, we analysed the dust-corrected luminosity of the [OIII]$\lambda$5007 line, $Lum[OIII]$. We find that, in counter-rotating hosts, the AGN classified as Composite and Ambiguous have a slightly enhanced nuclear activity. Nevertheless, the median of the $Lum[OIII]$ values is almost identical (within the estimated errors) for Seyfert and LINER galaxies. 

\end{itemize}

\section{Acknowledgments}
The authors thank Dr. Xavier Hernandez for a detailed revision and useful comments.

This work was partially supported by the Consejo Nacional de Investigaciones
Cient\'{\i}ficas y T\'ecnicas and the Secretar\'{\i}a de Ciencia y T\'ecnica 
de la Universidad Nacional de San Juan.

Funding for the SDSS has been provided by the Alfred P. Sloan
Foundation, the Participating Institutions, the National Science Foundation,
the U.S. Department of Energy, the National Aeronautics and Space
Administration, the Japanese Monbukagakusho, the Max Planck Society, and the
Higher Education Funding Council for England. The SDSS Web Site is
http://www.sdss.org/.

The SDSS is managed by the Astrophysical Research Consortium for the
Participating Institutions. The Participating Institutions are the American
Museum of Natural History, Astrophysical Institute Potsdam, University of
Basel, University of Cambridge, Case Western Reserve University,
University of
Chicago, Drexel University, Fermilab, the Institute for Advanced Study, the
Japan Participation Group, Johns Hopkins University, the Joint Institute for
Nuclear Astrophysics, the Kavli Institute for Particle Astrophysics and
Cosmology, the Korean Scientist Group, the Chinese Academy of Sciences
(LAMOST), Los Alamos National Laboratory, the Max-Planck-Institute for
Astronomy (MPIA), the Max-Planck-Institute for Astrophysics (MPA), New Mexico
State University, Ohio State University, University of Pittsburgh, University
of Portsmouth, Princeton University, the United States Naval Observatory, and
the University of Washington.

\label{lastpage}
\end{document}